\documentclass{elsart} 
\usepackage{epsfig,amssymb} 
\journal{Physica A}  

\begin{document} 

\begin{frontmatter}  

\title{Two-dimensional dissipative maps at chaos threshold: 
       Sensitivity to initial conditions and relaxation dynamics}
\author{Ernesto P. Borges}
 \ead{ernesto@ufba.br}
 \address{Escola Polit\'ecnica, Universidade Federal da Bahia, \\
          Rua Aristides Novis 2, 40210-630 Salvador-BA, Brazil 
          \\ \and 
          Centro Brasileiro de Pesquisas F\'\i sicas \\
          R. Dr. Xavier Sigaud, 150, 22290-180 Rio de Janeiro-RJ, Brazil
         }
\author{Ugur Tirnakli\corauthref{cor}}  
 \corauth[cor]{Corresponding author.} 
 \ead{tirnakli@sci.ege.edu.tr} 
 \address{Department of Physics, Faculty of Science, Ege University, 
          35100 Izmir, Turkey}   

\begin{abstract} 
The sensitivity to initial conditions and relaxation dynamics of 
two-dimensional maps are analyzed at the edge of chaos, along the lines of 
nonextensive statistical mechanics. 
We verify the dual nature of the entropic index for the Henon map,
one ($q_{sen}<1$) related to its sensitivity to initial conditions properties,
and the other, graining-dependent ($q_{rel}(W)>1$), related to its 
relaxation dynamics towards its stationary state attractor.
We also corroborate a scaling law between these two indexes, previously found
for $z$-logistic maps.
Finally we perform a preliminary analysis of a linearized version of the
Henon map (the smoothed Lozi map). We find that the sensitivity properties
of all these $z$-logistic, Henon and Lozi maps are the same, 
$q_{sen}=0.2445\dots$
\end{abstract}  

\begin{keyword} 
Nonextensive thermostatistics
\sep
dynamical systems
\sep
two-dimensional maps

\PACS 05.45.-a \sep 05.45.Ac \sep 05.20.-y
\end{keyword} 

\end{frontmatter}  

\date{\today} 
\maketitle   

\section{Introduction} 
The rebirth of the focus on nonlinear maps
has opened unsuspected new perspectives in physics. 
Among them, the logistic map \cite{feigenbaum1,feigenbaum2}
has achieved a paradigmatic rank. But not only dissipative maps have been 
playing important roles; also conservative and symplectic ones,
like the standard map, have their places.
These simple nonlinear models exhibit essential features of ordered and 
chaotic systems, and also of the transition between them.
When we deal exactly with the critical points, these models present
many typical behaviors of complex systems, for instance,
slow dynamics, break of ergodicity, power-laws, self-similarity, among others.	

One of the important contributions that comes from the study of these maps 
aims to
understand the fundamental connections between dynamics and
thermodynamics. In fact, the relation between these two main branches
of physics is not yet fully understood and there are many points that
deserve further investigations \cite{cohen}.
Once these systems have low-dimensionality, it is possible to computationally
investigate them in its full phase space, something that is very hard 
(not to say impossible) to implement for high-dimensional systems.  
There are also important analytical results regarding the logistic maps
at the edge of chaos \cite{fulvio-alberto:1,fulvio-alberto:2}.

It is very impressive that features exhibited by high-dimensional Hamiltonian
systems \cite{vito-andrea-ct} may be found in some of these maps 
(conservative \cite{fulvio-edgardo-ct} 
or even dissipative \cite{fulvio-luis-ana-alberto-ct}), e.g.,
the meta-stationary non-Boltzmannian state with a crossover 
to a stationary Boltzmannian state 
(crossover that is shifted in time as the system grows), indicating that the 
scenario conjectured in \cite{ct:bjp} is somehow ubiquitous.

There are basically two computational paths for investigating such 
dissipative
systems, along nonextensive lines: sensitivity-based and relaxation-based 
approaches.
By {\em nonextensive} we mean the use of the nonextensive entropy
(written in its dimensionless form)
\vspace{-0.5cm}
\begin{eqnarray}
 S_q=\frac{1-\sum_{i=1}^W p_i^q}{q-1} \qquad (q \in \mathbb{R}),
\end{eqnarray}
and the concepts it implies.
The first step of both approaches is to define a partition of the phase space 
into $W$ equally spaced cells (the graining). Then the graining is made 
finer (by increasing $W$).
In the sensitivity-based approach, we choose one of the $W$ cells of the 
phase space, put within it $N$ points (typically $N \gg W$), 
and then follow the dynamical evolution of these points. 
Its initial entropy is $S_q(t=0)=0$, $\forall q$, (complete knowledge) 
so it increases with time.
The chosen initial cell is taken as that (or those) that rapidly spread
the points over the phase space.
It is possible to estimate the value of the entropic index,
here called $q_{sen}$, by at least three different methods, all of them
giving the same results.
To see a description of these methods, when they were first employed, see 
\cite{ct-angel-zheng,marcelo-ct:prl,vito-baranger-andrea-ct}.

In the second (relaxation-based) approach \cite{moura-ugur-marcelo}, we 
initially take $N \gg W$ points uniformly distributed in the whole phase space. 
Its initial entropy $S_q(t=0) = (W^{1-q}-1)/(1-q)$ is a maximum, $\forall q$
(complete lack of information).
At the edge of chaos and as $t\to\infty$, the distribution of points do not 
occupy the whole phase space, but a fractal subspace of it. 
So the entropy decreases (relaxes) with time. The shrinking of the occupied volume 
permits to estimate the proper entropy index, here called $q_{rel}$.

These two approaches to the same problem generate different entropic
indexes: the sensitivity-based approach generates $q_{sen}\le 1$,
while the relaxation-based approach generates $q_{rel}\ge 1$.
In Ref. \cite{epb-ct-garin-murilinho} it was found for the logistic map
that the entropic index $q_{rel}$ depends on the graining $W$, 
and that there exists the following scaling-law relation:
\vspace{-0.5cm}
\begin{eqnarray}
 \label{eq:scaling}
 q_{rel}(W\to\infty) - q_{rel}(W) \propto W^{-|q_{sen}|}.
\end{eqnarray}
In the present work we further investigate
this scaling relation for the Henon map (preliminary results were given 
in \cite{epb-ugur:physicaD}).
The existence of a scaling-law between sensitivity-based and relaxation
(graining dependent)-based approaches for different systems, including
high-dimensional ones, was suggested in \cite{ct-epb-fulvio:cagliari2001}.
We also perform a preliminary study of a linearized version of the Henon map,
according to the sensitivity-based approach.

Before going on, let us make a few comments about nomenclature. 
In early papers on nonextensive statistical mechanics, there were no
awareness of the duality of the $q$ indexes, so they were simply called 
$q$, or, sometimes, $q^*$. 
The existence of the duality was shown in \cite{moura-ugur-marcelo} 
and then commented in \cite{ct:duality}.
In that paper it was adopted the expression {\em mixing} in the sense of 
sensitivity to initial conditions, and {\em equilibrium} in the sense of 
relaxation to the stationary state attractor.
So it was then employed the nomenclature $q_{mix}$ and $q_{eq}$. 
In some later papers on the subject, this nomenclature has gradually 
changed. 
In the present work we adopt $q_{sen}$ in the place of $q_{mix}$, 
and $q_{rel}$ in the place of $q_{eq}$, due to the generality of the concepts.
Such adaptation of concepts and nomenclature is natural in a 
still evolving branch of science.

\section{Sensitivity and relaxation for two-dimensional maps}

We analyze two examples of two-dimensional maps. 
The first one is the Henon map \cite{henon}:
\vspace{-0.5cm}
\begin{eqnarray}
 \label{eq:henon}
 \left\{
  \begin{array}{lll}
   x_{t+1} &=& 1 - a x_t^2 + y_t, \\
   y_{t+1} &=& b x_t.
  \end{array}
 \right.
\end{eqnarray}
The parameter $b$ recovers the logistic map when it is set to zero.
If $b=1$, the Henon map is conservative.
We are particularly interested at small values of $b$ (dissipative systems)
at the edge of chaos (critical points at which the standard Lyapunov exponent 
changes sign).
The Table \ref{tab:henon} presents the critical values of the parameter $a$
for different values of $b$, as well as the limits of the phase space in 
each case.
\begin{table}[htb!]
 \begin{center}
 \caption{Critical parameter $a=a_c$ for the Henon map, at the edge of chaos.
          $x_{min}$, $x_{max}$, $y_{min}$ and $y_{max}$ represent the limits
	  of the phase space along $x$ and $y$ axis, respectively.\medskip}
 \label{tab:henon}
 \begin{tabular}{c|c|c|c|c|c}
  \hline
  $b$  &   $a_c$          & $x_{min}$ & $x_{max}$ & $y_{min}$ & $y_{max}$ \\
  \hline
  0     & 1.40115518\dots     & -1     & 1     & 0         & 0      \\
  0.001 & \quad1.39966671\dots  & -1.05  & 1.05  &   -0.0015 & 0.0015 \\
  0.01  & 1.38637288\dots  & -1.05  & 1.05  &   -0.015  & 0.015  \\
  0.1   & \quad1.26359565\dots & -1.1   & 1.1   &   -0.10   & 0.12   \\
  0.2   & 1.14904648\dots   & -1.179 & 1.179 &   -0.236  & 0.236  \\
 \hline 
\end{tabular}
\end{center}
\end{table}

We also analyze the approximation of the Lozi map made at Ref. \cite{grebogi}. 
The Lozi map is a piecewise linearization of the Henon map. In \cite{grebogi}
the authors have made a smoothness in the original Lozi formulation
by inclusion of a third parameter $\epsilon$:
\vspace{-0.5cm}
\begin{eqnarray}
 \label{eq:lozi}
 \left\{
  \begin{array}{lll}
   x_{t+1} &=& 1 - a F_{\epsilon}(x) + y_t, \\
   y_{t+1} &=& b x_t,
  \end{array}
 \right.
\end{eqnarray}
and $F_{\epsilon}(x)$ is defined as
\vspace{-0.5cm}
\begin{eqnarray}
 F_{\epsilon} = 
 \left\{
 \begin{array}{ll}
  |x|                                          & \quad\mbox{if }|x|\ge\epsilon, 
  \\
  g_{\epsilon}(x)=(x^2/2\epsilon)+(\epsilon/2) & \quad\mbox{if }|x|\le\epsilon,
 \end{array}
 \right.
\end{eqnarray}
with $0<\epsilon<1$. The limit $\lim_{\epsilon\to0^+}F_{\epsilon}(x)=|x|$
recovers the original Lozi map.

\begin{figure}[htb]
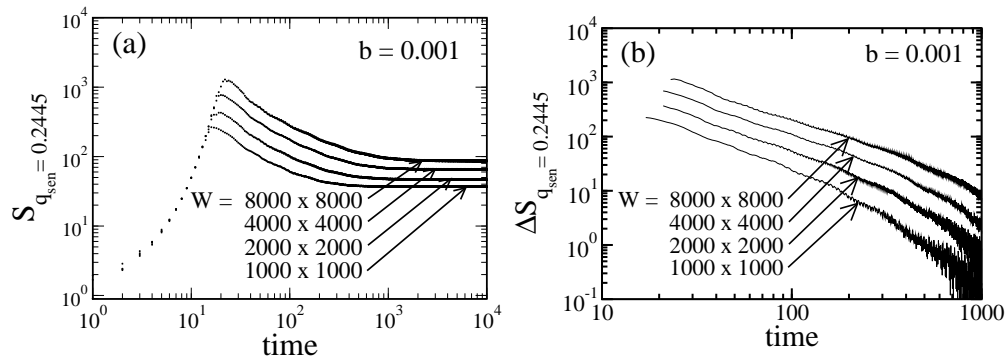

 \begin{minipage}[htb]{0.47\textwidth}
 \begin{center}
  \epsfig{file=borges-fig1a.eps,width=\textwidth,clip=}
 \end{center}
 \end{minipage}
 \begin{minipage}[htb]{0.47\textwidth}
 \begin{center}
  \epsfig{file=borges-fig1b.eps,width=\textwidth,clip=}
 \end{center}
 \end{minipage}
 \caption{(a) Time evolution of the $q_{sen}$-entropy for the Henon map;
          (b) Time evolution of $\Delta S_{q_{sen}}$.
          Different values of $W$, with $b=0.001$, at the edge of chaos.}
 \label{fig:Sq-vs-time}
\end{figure}

In order to establish a connection between $q_{sen}$ and $q_{rel}$,
we follow the procedure adopted in Ref. \cite{epb-ct-garin-murilinho},
that we briefly describe in the following:
we begin with $N$ points inside a properly chosen single cell (similar to the 
sensitivity-based approach). The initial cell is chosen as that one which
gives the maximum spread of the points --- a maximum overshoot in the
curve $S_{q_{sen}}$ vs. time (see Fig. \ref{fig:Sq-vs-time}(a), illustrated
with the Henon map).
The entropy is evaluated with the value $q_{sen}$, as determined by
the sensitivity-based methods.
Fig. \ref{fig:Sq-vs-time}(a) may be divided into two parts: the ``uphill side''
is governed by the sensitivity to the initial conditions;
the ``downhill side'' is governed by the relaxation to the final
stationary value of $S_{q_{sen}}$.
These ``two sides'' are a visual expression of the non-commutability
of the limits $\lim_{t\to\infty}\;\lim_{W\to\infty}$ and
$\lim_{W\to\infty}\;\lim_{t\to\infty}$. The former expresses that
complete knowledge never meets a stationary state. On the other hand,
the later order of limits expresses that a stationary state is only 
possible with lack of information. This non-commutability was conjectured
in Ref. \cite{ct:bjp} for many-body Hamiltonian systems, if we substitute
the present variable $W$ for the number of particles.

We particularly focus on the ``downhill side'' and define the variable
$\Delta S_{q_{sen}} \equiv S_{q_{sen}}(t) - S_{q_{sen}}(t\to\infty)$.
The time evolution of $\Delta S_{q_{sen}}$ gradually displays a
power-law, as $W$ (and proportionally $N$) increases 
(Fig. \ref{fig:Sq-vs-time}(b)).
The slope (in a log-log plot) of the curve, in the range $\Delta S_{q_{sen}}$ 
follows a power-law, is identified with the power of a decreasing 
$q$-exponential (without cutoff), 
$e_q(-t)=[1-(1-q_{rel})\,t/\tau_q]^{1/(1-q_{rel})}$
($|\mbox{slope}|=1/(q_{rel}(W)-1)$).
It is somehow subjective to decide when the power-law begins and when
it ends. So we average over many possible intervals (different beginnings
and endings), taking into account those with higher correlation coefficient.

\begin{figure}[htb]
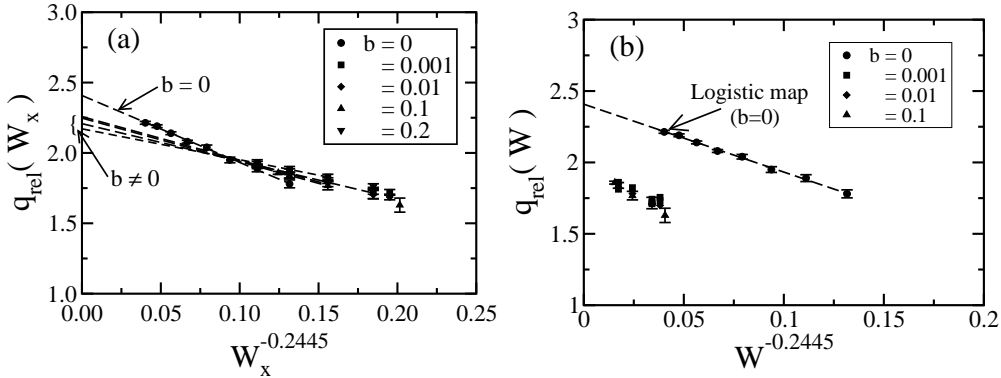

 \begin{minipage}[htb]{0.47\textwidth}
 \begin{center}
  \epsfig{file=borges-fig2a.eps,width=\textwidth,clip=}
 \end{center}
 \end{minipage}
 \begin{minipage}[htb]{0.47\textwidth}
  \begin{center}
   \epsfig{file=borges-fig2b.eps,width=\textwidth,clip=}
  \end{center}
 \end{minipage}
 \caption{Scaling law for the Henon map (Eq.~(\protect\ref{eq:scaling})),
          at the edge of chaos. 
	  (a) The abscissa is the partition of the logistic subspace, $W_x$.
	  (b) The abscissa is the full partition of the two-dimensional
	  phase space, $W$, as appeared in 
	  Ref. \protect\cite{epb-ugur:physicaD}.
	  The Figures make evident that the choice (b) is not fair, as 
	  explained in the text.
	  }
 \label{fig:scaling}
\end{figure}

The sensitivity-based approach to this map was applied previously 
\cite{ugur:pre}, when it was found $q_{sen}=0.2445\dots$, 
the same value found for the logistic map
\cite{ct-angel-zheng,marcelo-ct:prl,vito-baranger-andrea-ct}.
When the relaxation-based method is applied to the Henon map 
\cite[Fig. 1]{epb-ugur:physicaD}, once again it was found the same value of
the logistic map for infinite fine graining, $q_{rel}(W\to\infty)=2.41\dots$.
These results may include the Henon map into the family of the logistic map.
We then examine the relation between these two $q$ indexes:
we approximately find the scaling law, Eq. (\ref{eq:scaling}). 
The scaling law is found when we analyze the projection
of the map in its {\em logistic subspace} ($x$ axis), so we use $W_x$
instead of $W$ in Eq. (\ref{eq:scaling}).
Fig. \ref{fig:scaling}(a) shows our results for the Henon map.
The errors associated with the obtained results are relatively 
large when compared with those of the logistic map. 
This may be due to the fact that two-dimensional maps demand higher number
of cells $W$, and consequently higher number of points $N$. Once the computer 
capabilities are, of course, limited, the graining we can achieve in 
two-dimensional maps are coarser than in one-dimensional ones.
For this reason, the extrapolated curves for $q_{rel}(W\to\infty)$ in 
Fig. \ref{fig:scaling}(a) for $b\ne 0$ should be viewed as indicative.
We also conclude that the logistic limit $b\to 0$ is slowly reached.
We call attention that in Ref. \cite{epb-ugur:physicaD} we were not able 
(at that moment) to see that it was necessary to analyze the logistic subspace,
once we had only one value of $b$; 
in that paper we adopted $W$ in the scaling law. 
Only now we amend that conclusion: Fig. \ref{fig:scaling}(b) shows that 
the use of the full partition of the phase space $W$, in the two-dimensional 
Henon map is not fair, once $b\to 0$ does not recover the logistic case.

\begin{figure}[htb]
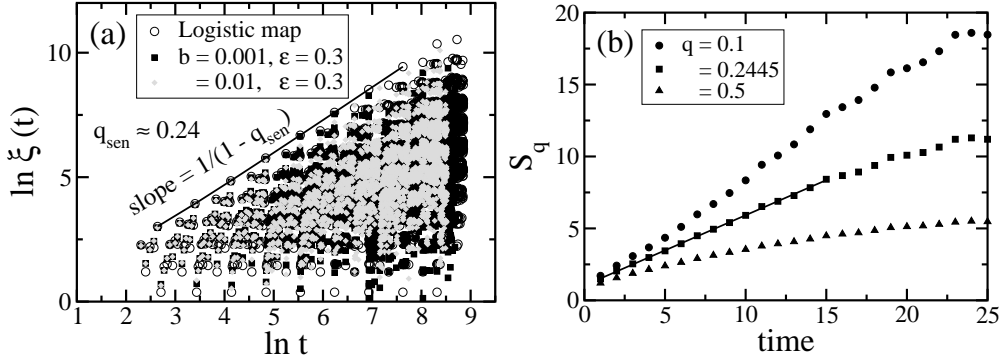

 \begin{minipage}[htb]{0.47\textwidth}
 \begin{center}
  \epsfig{figure=borges-fig3a.eps,width=\textwidth,clip=}
 \end{center}
 \end{minipage}
 \begin{minipage}[htb]{0.47\textwidth}
  \begin{center}
   \epsfig{figure=borges-fig3b.eps,width=\textwidth,clip=}
  \end{center}
 \end{minipage}
 \caption{Smoothed Lozi map (Eq. (\protect\ref{eq:lozi})), with $\epsilon=0.3$.
	  (a) Time evolution of the sensitivity to initial conditions.
          (b) Time evolution of $S_q$.	  }
 \label{fig:lozi}
\end{figure}

Let us now address our results for the smoothed Lozi map, Eq. (\ref{eq:lozi}),
at the edge of chaos (when $\epsilon=0.3$, $a_c=1.72879898\dots$ for $b=0.001$, 
and $a_c=1.71996303\dots$ for $b=0.01$).
We have focused this map according to the sensitivity-based
approach. The analysis of the sensitivity to initial conditions at the
edge of chaos (see \cite{ct-angel-zheng} for details of this method) 
shows a vanishing standard Lyapunov exponent. The positive $q$-Lyapunov 
parameter, $\lambda_{q_{sen}}$, expresses that the system follows a 
complex dynamics, whose upper bound of the sensitivity $\xi(t)$ evolves 
through a slow dynamics, 
$\xi(t)=[1+(1-(q_{sen})\,\lambda_{q_{sen}}\,t]^{1/(1-q_{sen})}$,
with $\xi(t)\equiv\lim_{\Delta x(0)\to0}\frac{\Delta x(t)}{\Delta x(0)}$.
Fig. \ref{fig:lozi}(a) identifies the slope of the upper bound with
the power of the previous $q$-exponential equation for $\xi(t)$.
A different method related to this sensitivity-based approach, namely the
rate of increase of $S_q$ \cite{vito-baranger-andrea-ct}, is also presented
(Fig. \ref{fig:lozi}(b)). 
The proper value of $q$ is that one which generates a linear increase of 
entropy (a {\em finite} rate of entropy production).
Both methods give $q_{sen}\approx0.2445$, the same value of the logistic map.
It is still missing the analysis of this smoothed Lozi map
according to the relaxation-based method, as well as the correlation
between its entropic indexes $q_{sen}$ and $q_{rel}$.

\section{Conclusions}  
The duality among the $q$-indexes, one related to approaches based on
sensitivity to initial conditions, $q_{sen}$, and other related to
approaches based on relaxation to stationary states, $q_{rel}$,
and the connection between them,
being verified for the first time for the logistic map 
(Ref.  \cite{epb-ct-garin-murilinho}), has been approximately verified in a 
two-dimensional (Henon) map, which belongs to the logistic family.
The relation is verified in the logistic subspace of the two-dimensional map.
The results straighten the validity of the scaling relation,
Eq. (\ref{eq:scaling}). 
Possibly similar scaling laws exist for different phenomena, 
as suggested in Ref. \cite{ct-epb-fulvio:cagliari2001}, and further
studies along these lines would be very welcome.
The preliminary analysis of the smoothed Lozi map,
that introduces a linearization in the Henon map, indicates that,
as far as the sensitivity to initial conditions are concerned, 
this map presents the same value of $q_{sen}$ as that of the logistic and 
Henon map.
Further studies are necessary to evaluate the characteristic values
$q_{rel}$, as well as the connection between them.
Recently it has been conjectured in \cite{ct:physicaD-LosAlamos} that the 
complete characterization of a system may demand not only two, but 
{\em three} entropic indexes. The third one would be related to the 
stationary state itself, named $q_{stat}$.
These indexes would certainly be related among themselves. 
Detection and estimation of $q_{stat}$, as well as the discovery of how these 
three indexes are connected, are essential steps towards the closure of the 
nonextensive statistical mechanics theory.

\section*{Acknowledgments} 
\vspace*{-0.5cm}
This work has been partially supported by the Turkish Academy of Sciences, 
in the framework of the Young Scientist Award Program (UT/TUBA-GEBIP/2001-2-20).
Fulvio Baldovin is acknowledged for his remarks.
The authors gratefully acknowledge the hospitality of the organizers
of the Second Sardinian's NEXT 2003 meeting.


\end{document}